\newcommand{\CLASSINPUTinnersidemargin}{1in}
\newcommand{\CLASSINPUToutersidemargin}{1in}
\newcommand{\CLASSINPUTtoptextmargin}{0.76in}
\newcommand{\CLASSINPUTbottomtextmargin}{1in}

\documentclass[11pt,conference,letterpaper]{IEEEtran}
\IEEEoverridecommandlockouts
\usepackage{cite}
\usepackage{amsmath,amssymb,amsfonts}
\usepackage[ruled,vlined]{algorithm2e}
\usepackage{graphicx}
\usepackage{textcomp}
\usepackage{xcolor}
\usepackage{mathtools}
\usepackage{booktabs}
\usepackage{multirow}
\usepackage{multicol}
\usepackage{tabularx}
\usepackage{adjustbox}
\usepackage{rotating}
\usepackage{url}
\usepackage{enumitem}
\def\BibTeX{{\rm B\kern-.05em{\sc i\kern-.025em b}\kern-.08em
    T\kern-.1667em\lower.7ex\hbox{E}\kern-.125emX}}
\begin{document}
\pdfoutput=1
\title{User Allocation in Heterogeneous Network Supplied by Renewable Energy Sources\\
\thanks{The work has been funded within the statutory expenditures — project no. 0312/SBAD/8160}}

\author{\IEEEauthorblockN{Adam Samorzewski}
\IEEEauthorblockA{\textit{Institute of Radiocommunications} \\
\textit{Poznan University of Technology}\\
Poznan, Poland \\
adam.samorzewski@student.put.poznan.pl}
\and
\IEEEauthorblockN{Adrian Kliks}
\IEEEauthorblockA{\textit{Institute of Radiocommunications} \\
\textit{Poznan University of Technology}\\
Poznan, Poland \\
adrian.kliks@put.poznan.pl}
}

\maketitle

\begin{abstract}
The information and communication technology, ICT, a domain of the global economy, consumes more and more energy, what is the effect of continuously increasing wireless data traffic. Thus, there is a need for more energy-efficient and sustainable network and resource management, and such a goal can be achieved by the utilization of renewable energy sources. This paper proposes a novel user allocation scheme applicable to the heterogeneous radio access networks supplied with renewable energy sources. Presented results proved that high-energy efficiency can be accomplished by reallocating some traffic to the green wireless base station\footnote{Copyright © 2021 IEEE. Personal use is permitted. For any other purposes, permission must be obtained from the IEEE by emailing pubspermissions@ieee.org. This is the author’s version of an article that has been published in the proceedings of 2021 17th International Conference on Wireless and Mobile Computing, Networking and Communications (WiMob) and published by IEEE. Changes were made to this version by the publisher prior to publication, the final version of record is available~at: http://dx.doi.org/10.1109/WiMob52687.2021.9606428.~To~cite the paper use:~A.~Samorzewski and A.~Kliks,~“User Allocation in Heterogeneous Network Supplied by Renewable Energy Sources,” 2021 17th International Conference~on Wireless and Mobile Computing,~Networking and Communications (WiMob), 2021, pp.~419-422, doi: 10.1109/WiMob52687.2021.9606428 or visit https://ieeexplore.ieee.org/document/9606428}. 
\end{abstract}


\section{Introduction}
\label{sec_intro}

% From one year to another, the delivery of wireless services to mobile users is being improved in terms of offered network capacity, signal coverage, achieved throughput and reliability \cite{Boccardi2014,Yang2019,Kliks2019}.
Progressive standardization of radio access network (RAN) allows achieving more and more stringent requirements on the quality of service experienced by the end-user \cite{Saha2018}. As the number of served users is also increasing, the structure of the network becomes more and more heterogeneous, where many base stations of a different kind (macro- and micro- base stations, small cells, etc.) are deployed on the same area, leading to the concept of no-cell schemes \cite{Bjornson2020}. However, in such a complicated and demanding scenario, the reliable service delivery entails also the permanent increase in the energy consumption by the wireless part of the ICT sector of the global economy \cite{Lornicz2019}. As this problem is technically tractable in urban areas, where access to stable energy sources can be easily guaranteed, it may become a significant challenge in the rural, remote, and hardly accessible areas. The additional investment costs (mainly into the reliable power supply besides the telecommunication infrastructure) versus expected revenues from such areas can make such a project unprofitable. 
% However, despite the high capital expenditures related to remote-area investments, also the operational costs of the wireless nodes should be considered, regardless of the deployment area.
Moreover, the environmental perspective has recently gained great attention in the context of ICT - the communication network shall be both green (i.e. with reduced energy consumption) but also should target the resource-sustainability goal \cite{Lornicz2019,Fletscher2019}. This sustainability of resources is an important paradigm that can drive the development of new algorithms and solutions for wireless networks \cite{Hossain2020}.   
The three cases identified above (i.e., investment in hardly accessible areas, reduction of energy consumption, and improvement of sustainable usage of resources), are to some extent addressed by equipping the base stations with renewable energy sources (RESes), such as solar panels, wind turbines, etc. \cite{Fletscher2019}. Providing the portable base station with RESes (which make the base station to some extent energy-autonomous) could be a viable solution for the delivery of wireless services in remote areas. 
% Next, RES may lead to the reduction of the consumption of the energy delivered by traditional (grid) sources, improving the overall sustainability of resource usage \cite{Li2015,Jahid2019,Lam20200}. RESes collect energy from the surrounding environment (sun, wind, etc.) and accumulate it either in local batteries or deliver it back to the electricity grid (if available).
% Unfortunately, the RESes are strictly dependent on the actual weather conditions, thus - by nature - are or may be highly unstable. This in consequence can have a significant and negative impact on the quality of services experienced by mobile users. Thus, it is necessary to provide an additional backup energy source to guarantee smooth service delivery. In case when the electricity grid is not available (as in remote areas), the application of accumulators together with the diesel generators as the alternative source of energy could be a solution \cite{Fletscher2019}. 
However, the application of RESes entails the need for prospective adjustment or even redesign of various existing algorithms operating at the base station, such as radio resource management, user assignment, traffic steering, etc.
% In particular, the instability of RESes as well as the knowledge about the remaining energy stored in the batteries have to be considered. For example, it may be appropriate to re-associate users from a low-battery base station to the one where the battery is full. The problems of energy-aware networking as well as of energy-source instability have been widely talked about in the context of sensor networks, e.g. \cite{Xu2015}, however when dealing with high load broadband access system, new aspects have to be taken into consideration.
Thus, in this paper, inspired by the work in \cite{Fletscher2019}, we propose the solution of RES-aware user assignment procedure. In particular, we have proposed the solution, where the new user is allocated at first to green nodes; when it is impossible it tries to reallocate already assigned users among the green nodes in order to relax some resources to serve the new user. Only when these steps were not successful, the user is allocated to the on-grid macro base station. Moreover, the proposed solution has been extended to be applicable in two other schemes - user handover and enforced switching-off of the green node. 

\section{Scenario Description}
\label{sec_scenario}
\begin{figure}[ht]
\centerline{\includegraphics[width=0.47\textwidth]{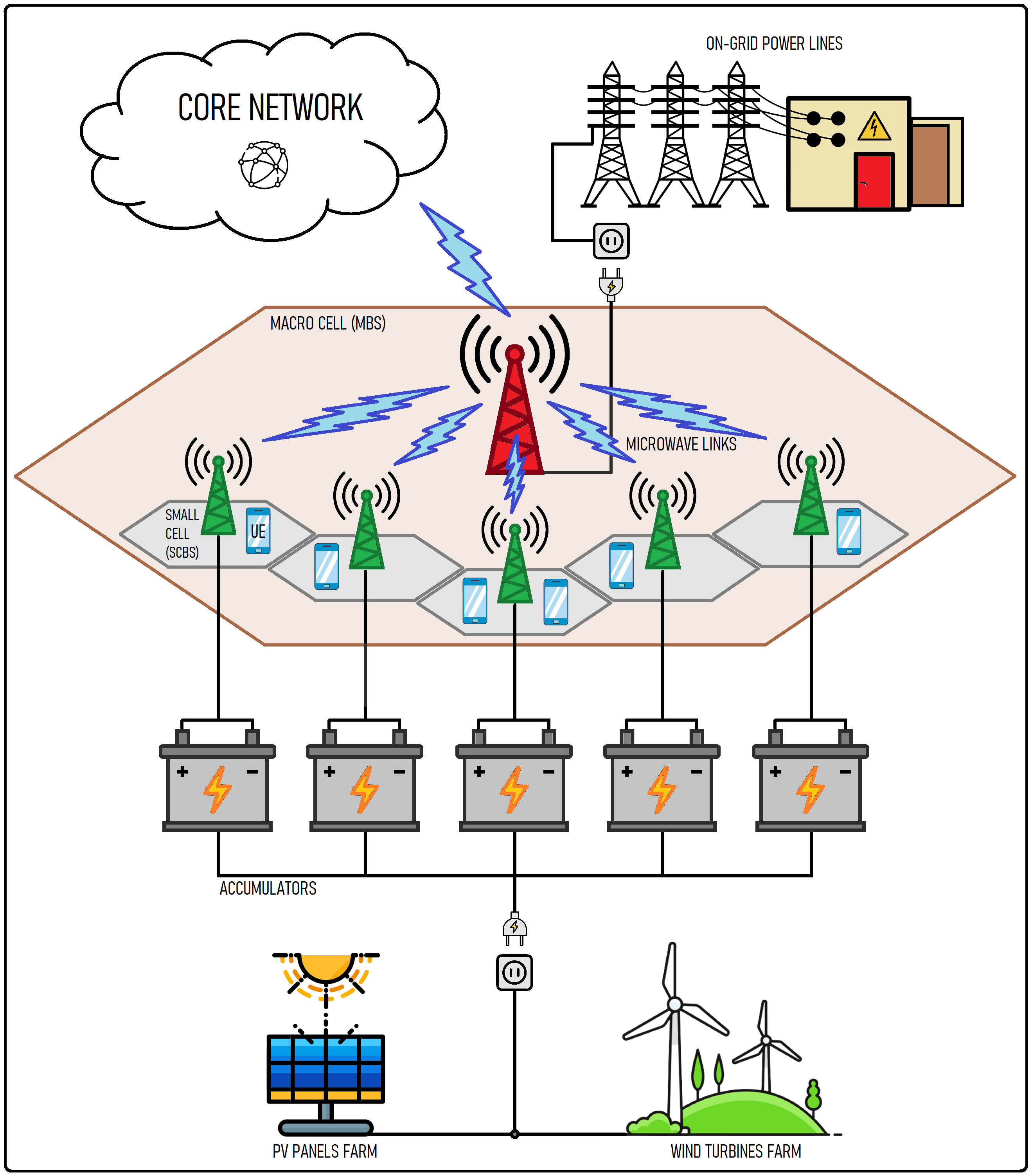}}
\caption{Considered scenario}
\label{fig_scenario}
\end{figure}

In our analysis, we consider the heterogeneous 5G network consisting of one macro base station (MBS) cell equipped with the on-grid power supplies, and $N-1$ small cell base stations (SCBSes) powered by renewable energy sources (Fig.~\ref{fig_scenario}).  The MBS is located in the center of the examined area and acts as the sink node (due to the fixed wireless access to the core network) to the surrounding green SCBSes constituting the star topology for the backhaul connections.
% Without loss of generality, the SCBSes are deployed uniformly over the considered area, creating a regular grid of fixed size. 
We assume that the cells can be sectorized, and per each sector, multiple antennas can be installed on the site. Mobile users are deployed randomly and uniformly in the considered area, and they can change their positions. Two classes of users are envisaged: the low-rate (which requires up to 3 resource blocks (RB) per one time slot) and the high-rate one (with 10 RBs per time slot). 
% The proposed user assignment algorithm is referred to the one presented in~\cite{Fletscher2019}.

We assume four sleep modes for SCBSes. Each subsequent mode is less and less energy-hungry at the expense of the increasing awaking time (followed by \cite{Salem2019}). In case of no users appearing for a longer time, SCBS enters the deeper sleep mode after a strictly defined time. 

As MBS is supplied by the on-grid sources, we can assume permanent power availability. In case of SCBS, we consider the photovoltaic (PV) panels and wind turbines (WT) grouped in farms as energy sources.
\section{RES-aware User Assignment}
\label{sec_assignment}
% Once the details on modelling of the power consumption, generation and storing are presented, we concentrate on the proposal of RESes aware user assignment scheme.
% As the reference scenario, we take the algorithm proposed in \cite{Fletscher2019}, which consists of few key steps and constraints. First, the users will be connected to the green node if it has enough spectral and energy resources to serve this user, and the decision on the availability of resources is part of the system design. However, if there's no RES supplied node, which meets certain requirements, an attempt is made to connect the user to the on-grid macro cell. But if the received signal from the MBS is too weak or it hasn't got enough spectral resources, the user will not be served and is considered to be in outage. %In this algorithm, the preference is given to the green SCBS, however we see the potential of more advanced user allocation procedure, where numerous SCBS and MBS are coordinated. 

In particular, as in the original algorithm, we prioritize the green SCBSes in the user assignment process. Thus, while associating the user with the serving base station, the first attempt is to find the best (with the strongest signal) SCBS with enough spectrum and energy resources to serve the user. In particular, we set the requirement that the amount of frequency resources will be at least enough to guarantee the rate required by the user. Moreover, the amount of energy remaining in the attached accumulator has to be above some specific threshold \textit{e}, considering the power generation and consumption models. In case that no SCBS can be identified, the algorithm tries to reallocate the users already associated with SCBSes among the whole set of SCBS. The purpose of the user re-association is to check the possibility to relax some of the currently utilized resources from the serving SCBS and allocate them to the new user. Only when this procedure did not lead to the positive user assignment of the user to green SCBS, the MBS is considered as the final candidate to serve this user. Finally, when there is no traffic generated within the SCBS cell, it enters the first sleep mode. If there is no traffic load for a specific time, the SCBS goes into deeper sleep mode, finally achieving the fourth mode. The SCBS will return to the normal operating mode when any new data traffic should be served. 
Let us stress that although the above procedure is directly applicable when the new user tries to access the network, it can be also launched periodically to update the user association to the base station in order to reflect the RESes instability and react to the movement of users. Thus, in our investigation, we assume that the user-to-base station association procedure will be triggered in the following situations: a user will initiate the handover procedure; any of the SCBS will announce about the insufficient energy stored in the batteries. Analogously, when the battery has been recharged and the amount of the stored energy has sufficiently increased above the threshold, the user re-association procedure can be also revoked.
\section{Simulation Results}
\label{sec_results}
\subsection{Simulation Setup}
To verify the performance of the proposed algorithm, extensive computer simulations using Matlab environment have been conducted, where the 5G NR base stations have been deployed as described in Sec.~\ref{sec_scenario}. There is one MBS was located centrally in the considered area, and 24 SCBSes deployed in form of a regular grid. The center frequency in FR1 band was set to 3410 MHz, and the channel bandwidth - to 40 MHz. It results in 106 resource blocks (RBs) available, assuming 30 kHz of subcarrier spacing. A Time Division Duplex, TDD, scheme with 2 uplink symbols located at the end of the slot was used. 

The MBS is a 3 sector site with a single antenna mounted at a height of 47 m, whose transmit power equals 46 dBm. In contrast, each SCBS is a 1 sector with a single antenna, 16 m height site, transmitting with the power of 32 dBm. In terms of backhaul connectivity, the microwave links have been used, in particular, there is a 24 MIMO array at the MBS, whereas at the SCBS - just one antenna is considered.

% The MBS is a 3 sector site with a 4-antenna MIMO array mounted at a height of 47 m, whose transmit power equals 46 dBm. In contrast, each SCBS is a 1 sector with an 8-antenna MIMO array, 16 m height site, transmitting with the power of 20 dBm. In consequence, it is considered, that there are 4 and 8 MIMO layers available for MBS and SCBS, respectively. In terms of backhaul connectivity, the microwave links have been used, in particular, there is a 24 MIMO array at the MBS, whereas at the SCBS - just one antenna is considered.

For the path loss calculation, the rural macrocell (RMa, \cite{3gppRMa}) has been considered with the following parameters: assumed slow fading margin - 6 dB, body loss - 3dB, foliage loss - 11 dB. The base station equivalent antenna gain was calculated to be 17.5 dB, whereas of the user - to 0dB. Thus, the resultant MBS coverage area has a radius of around 2.3 km, whereas the radius for SCBS was set to around 0.5 km. 
% In terms of applied energy efficiency parameters, the overhead power $P_{\mathrm{over}}$ per one sector was set to 6.8 W for SCBS \cite{Fletscher2019} and 130 W for MBS \cite{Guo2013}. The capacity threshold $\mathrm{Th}_{\mathrm{low}}$ was set to 2/3 of the allowable maximum capacity for both SCBS and MBS. The aggregated power for low and high traffic mode (i.e. $P_{\mathrm{low}}$ and $P_{\mathrm{high}}$) have been set to 12 and 30 W for SCBS, and to 37 and 92.5 W for MBS \cite{Monti2012}.
The MBS acts as the sink node, whose max aggregated backhaul capacity per sector was set to $27\times 106$  RBs, i.e. the sum of all RBs from 24 surrounding SCBSes and three sectors of the MBS.
% The single flow switch power was set to 53~W~\cite{Monti2012}.

% It is assumed that for single SCBS, there are 2 PV panels with an area of 2.25 m$^2$ and an efficiency ratio of $0.22$ each. Moreover, there is also a single WT with a rotor diameter of 2 m.
Assuming a moderately warm climate, the PV generates power on average 8 h per day. The energy is stored in one off-the-shelf accumulator model, i.e., Trojan 24-GEL 77Ah. There are 400 users (UE) deployed randomly with the uniform generator over the considered area. The UE antenna height is 1.5 m. Two classes of users are considered, the one that requires 3 RBs and the one that occupies 10 RBs. In both cases, the minimum acceptable signal-to-noise ratio is -2 dB. To calculate the approximate maximum data rate, we followed the guidelines provided in 5G R16 standard~\cite{3gppCapacity}.

% , i.e. 
% \begin{equation}
%     R_a = 10^{-6} \sum_{j=1}^J v_L^{(j)} Q_m^{(j)} f^{(j)} R_{\max} \frac{12N_{PRB}^{BW, (j),\mu}}{T_s^{\mu}} (1-OH^{(j)}),
% \end{equation}
% where $J$ stands for the number of aggregated component carriers, $R_{\max}$ = 948/1024. Per each carrier $j$, $v_L^{(j)}$ is the number of available layers, $Q_m^{(j)}$ - the maximum supported modulation order, $f^{(j)}$ - scaling factor, $\mu$ - 5G numerology, $T_s^{\mu}$ and $N_{PRB}^{BW, (j),\mu}$ are the average OFDM symbol duration and the maximum RB allocation in available bandwidth for given numerology. Finally, $OH^{(j)}$ is the applied overhead. In our system, besides the other parameters defined previously, we apply the overhead of XXXXX and scaling parameter equal to YYYYY. Moreover, the maximum Shannon capacity of the network has been simulated. Finally, in order to calculate the energy efficiency of the considered schemes, the following metric has been used, $EE = \frac{\sum_{n=1}^{N} E_{l=1}^{L} R_{n,l}}{\sum_{n=1}^{N} E_{l=1}^{L} E_{n,l}}$, i.e. the ratio of the sum rate and sum consumed energy in one second. $R_{n,l}$ and $E_{n,l}$ are the rate and consumed energy for transmission realized between base station $n$ and user $l$.

\subsection{User Assignment Analysis}
\begin{figure}[htb]
\centerline{\includegraphics[width=0.48\textwidth]{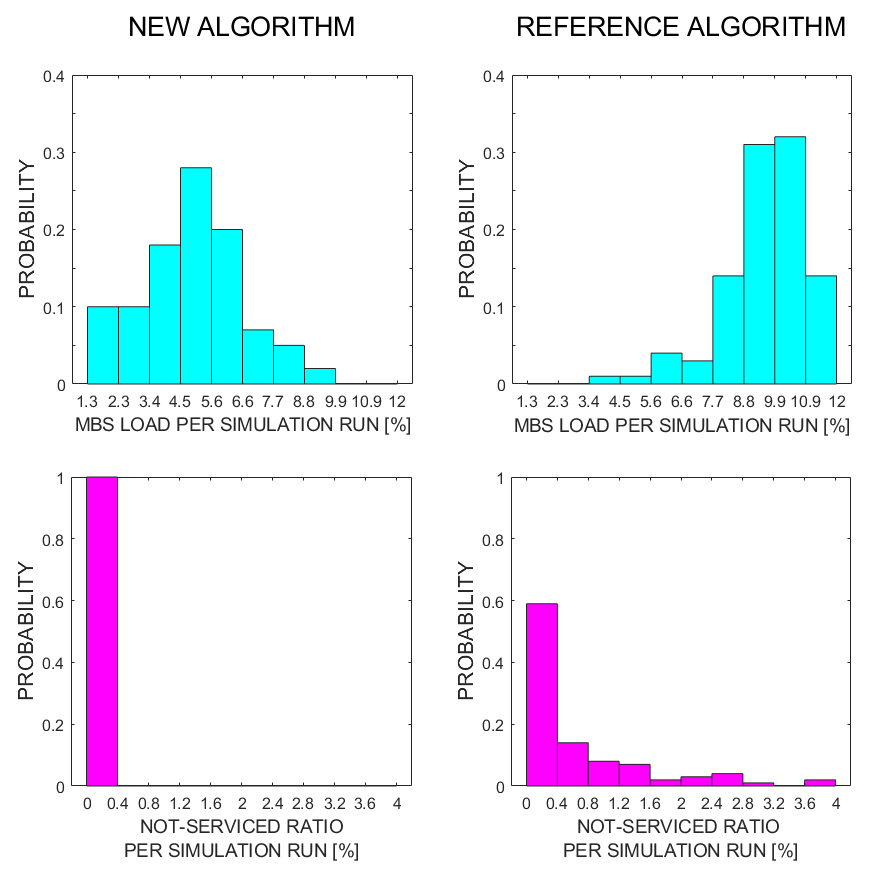}}
\caption{Assignment algorithms comparison.}
\label{fig_user_allocation}
\end{figure}

In the first step, we compare the distributions of the user allocation to base stations. The results presented in Fig.~\ref{fig_user_allocation} are achieved for 100 simulation runs with various user positions. The left subfigures show the results for the new algorithm, whereas the figures on the right - for the reference scenario. The first two top subfigures illustrate the distribution of the percentage load of the network served by the MBS. One can observe that for the new algorithm, the MBS processes on average only 5\% of the total traffic in the whole network. For the reference algorithm, the MBS served on average almost twice more traffic. Thus, a significant reduction of the MBS load has been observed. Moreover, in terms of the number of non-served users, the new algorithms have significantly reduced the outage probability. There were on average only 0.2\% of non-served users for the new algorithm.
%On the left side, there are the graphs represents new proposal of allocation algorithm, and the right side shows the results for reference algorithm. Graphs at the top describe percentage load of macro cell relative to total users number, and its incidence also as a percentage. In turn, the bottom histograms represents percentage of not-assigned users number, also relative to the total number, and its percentage incidence. By analyzing the charts, it can be concluded that new proposed solution of users allocation seems to be better than known reference algorithm. Arrangement of the histograms bins of reference oscillates around value of $10 \text{ } \%$ for MBS load, where for the new algorithm this value is about $5 \text{ } \%$. As it can be seen for new algorithm percentage of not-assigned users is also lower. The value is about $0 \text{ } \%$ for all simulation cases, but for reference solution not-assignment factor is more spread out, reaching values as high as around $4 \text{ } \%$.\\
\subsection{Continuous-time Analysis}
Let us now analyze the performance of the proposed algorithm in the continuous operation mode, i.e., when the network parameters are analyzed over the 72-hour-long period. 400 users have moved around along a pseudo-random trajectory, and the position of users have been changed around every 500s. All 24 SCBSes have at start fully charged accumulators. The network status has been stored every second to achieve reliable results. The proposed algorithm has been used at the simulation start for each user, when the user started handover procedure or when one of the BS was forced to be switched off due to the lack of energy from accumulator.

%AS comment - start (28.05)

% First, let us compare the distribution of the percentage of energy consumption per MBS and SBSC (top and bottom part of Fig.~\ref{fig_energy}, respectively) for new and reference algorithms (left and right subfigures). One can observe the shift of consumed energy from MBS to SCBS, thus better usage of energy from RES. 

% \begin{figure}[ht]
% \centerline{\includegraphics[width=0.5\textwidth]{updated_pictures/part-energy_popr.png}}
% \caption{Nodes energy consumption, divided by cell type, as a part of total network consumption.}
% \label{fig_energy}
% \end{figure}

%AS comment - end (28.05)

\begin{figure}[ht]
\centerline{\includegraphics[width=0.48\textwidth]{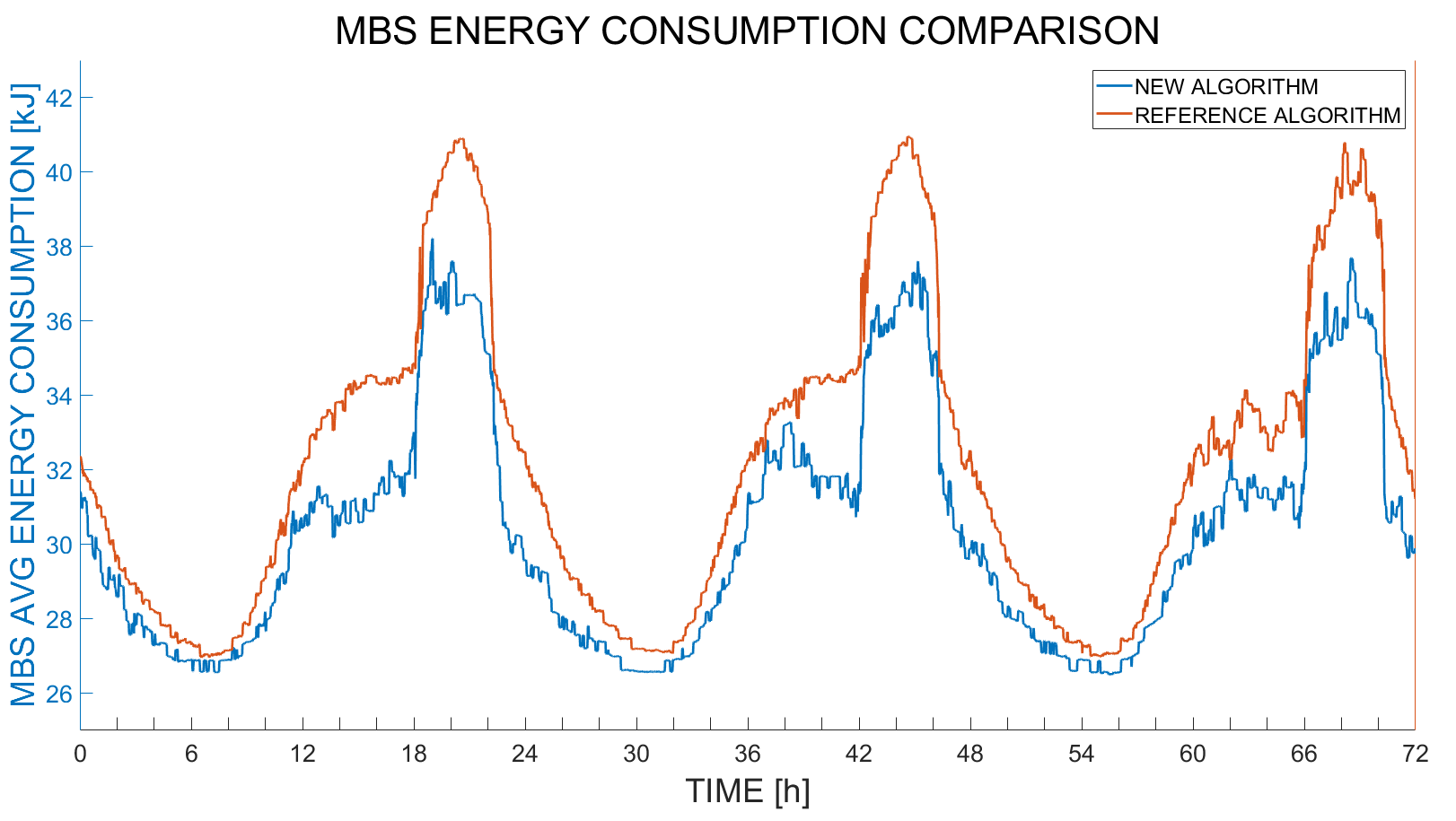}}
\caption{On-grid energy consumption comparison for both algorithms.}
\label{fig_time}
\end{figure}

%As the first one, Fig. 2 determines total network energy usage by resources division into amounts consumed by off-grid small cell and by on-grid macro cell. It can be seen superficially, that for new algorithm, the usage of energy resources is more concentrated around servicing by small cells, than for reference case. There was also observed, that for the new proposal of users assignment, on-grid resources make up smaller proportion of total energy consumption. More details are presented at the next figure.
As a remarkable gain in the usage of RES-based energy has been observed, it is worth analyzing the reduction of the on-grid energy consumption as a function of time, as illustrated in Fig.~\ref{fig_time}. One may observe a slight reduction of energy consumed by MBS, however, it has to be remembered that MBS is always powered on and it acts as the sink node. Thus, the prospective reduction of energy consumption of MBS is limited. 
Finally, we have compared the energy efficiency ratio, \textit{EE}, defined as the achieved sum rate of user data in the whole network over the total consumed energy averaged over all simulation steps. The results are presented in Tab.~\ref{tab_ee}, where the energy efficiency for SCBSes, MBS, and the whole network is jointly presented. One can observe that the energy efficiency has gained for green SCBSes (the difference was over 0.1 Mb per Joule per second), whereas there was a degradation of around 0.14 MbpJ/s per MBS. It is expected behavior, as the MBS is working as a sink node and has reduced the served traffic. Overall, the energy efficiency of the whole network was improved by around 0.04 Mb per Joule per second. 
\begin{table}[ht]
    \caption{Achieved avg energy efficiency in MbpJ per second}
    \label{tab_ee}
    \centering
    %\resizebox{0.5\textwidth}{!}{%
    \begin{tabular}{|p{1.5cm}|p{1.5cm}|p{1.5cm}|p{1.5cm}|} 
        \hline
        Algorithm & $EE_{\mathrm{SCBS}}$ & $EE_{\mathrm{MBS}}$ & $EE_{\mathrm{total}}$ \\
        \hline
        Reference & 1.75 & 0.39 & 1.25  \\ 
        \hline
        Proposed & 1.86 & 0.25 & 1.29  \\ 
        \hline
    \end{tabular}
    %}
\end{table}

\section{Conclusions}
\label{sec_concl}

In this paper, we have presented the new user-allocation scheme that is RES-aware, i.e., it tries to prioritize these base stations in the heterogeneous environment, which have access to renewable energy sources. Achieved results for static and dynamic (with user mobility) scenarios proved that the energy efficiency can be improved by shifting traffic from on-grid to RESes based nodes. However, the key problem in such an approach is the instability of RESes, and the resultant need for reliable coverage guaranteed by the on-grid base station. The latter cannot be switched off and enters the sleep mode more rarely than the small cells, thus the impact of the overhead power on overall power consumption is significant. 

\bibliographystyle{ieeetr}
\bibliography{references}

% \begin{thebibliography}{00}
% \bibitem{b1} G. Eason, B. Noble, and I. N. Sneddon, ``On certain integrals of Lipschitz-Hankel type involving products of Bessel functions,'' Phil. Trans. Roy. Soc. London, vol. A247, pp. 529--551, April 1955.
% \bibitem{b2} J. Clerk Maxwell, A Treatise on Electricity and Magnetism, 3rd ed., vol. 2. Oxford: Clarendon, 1892, pp.68--73.
% \bibitem{b3} I. S. Jacobs and C. P. Bean, ``Fine particles, thin films and exchange anisotropy,'' in Magnetism, vol. III, G. T. Rado and H. Suhl, Eds. New York: Academic, 1963, pp. 271--350.
% \bibitem{b4} K. Elissa, ``Title of paper if known,'' unpublished.
% \bibitem{b5} R. Nicole, ``Title of paper with only first word capitalized,'' J. Name Stand. Abbrev., in press.
% \bibitem{b6} Y. Yorozu, M. Hirano, K. Oka, and Y. Tagawa, ``Electron spectroscopy studies on magneto-optical media and plastic substrate interface,'' IEEE Transl. J. Magn. Japan, vol. 2, pp. 740--741, August 1987 [Digests 9th Annual Conf. Magnetics Japan, p. 301, 1982].
% \bibitem{b7} M. Young, The Technical Writer's Handbook. Mill Valley, CA: University Science, 1989.
% \end{thebibliography}
% \vspace{12pt}
% \color{red}
% IEEE conference templates contain guidance text for composing and formatting conference papers. Please ensure that all template text is removed from your conference paper prior to submission to the conference. Failure to remove the template text from your paper may result in your paper not being published.

\end{document}